\documentclass[aps,pra,groupedaddress,notitlepage, showkeys]{revtex4-1}

\usepackage{amssymb}
\usepackage{amsmath}
\usepackage{amsfonts}
\usepackage{amsthm}
\usepackage{graphicx}
\usepackage{xcolor}
\usepackage{bm}
\usepackage{url}
\usepackage{hyperref}
\usepackage{cases}

\begin{document}

\title{Structural and Lagrangian properties of analogue ensembles to characterize multifractality of stochastic processes}

\author{Carlos Granero-Belinch\'on} 
\affiliation{Department of Mathematical and Electrical Engineering, IMT Atlantique, Lab-STICC, UMR CNRS 6285, 655 Av. du Technop\^ole, Plouzan\'e, 29280, Bretagne, France.}
\affiliation{Odyssey, Inria/IMT Atlantique, 263 Av. G\'en\'eral Leclerc, Rennes, 35042, Bretagne, France.}
\email{carlos.granero-belinchon@imt-atlantique.fr}

\date{\today}

\begin{abstract}
We present a framework for the scale-invariance characterization of stochastic processes in reconstructed finite-dimensional spaces. This framework analyses the structural and dynamical properties of the reconstructed space and is based on a Takens embedding reconstruction followed by the definition of ensembles of analogue states. First, we specify a collection of target states densely sampling the full embedding space. For each target state, we search for the ensemble of its $k$-nearest neighbors, its $k$-best analogues, and we analyze its volume and dynamics. On the one hand, we study the probability distribution of the volumes and relate its mean and variance to the scale-invariance properties of the stochastic process. On the second hand, we study the Lagrangian properties of the analogues by characterizing how they disperse in time. More particularly, we study the volume occupied by the analogue's successors in function of time and of their initial volume. The main contribution of this paper is to highlight the link between these dynamical properties and the scale-invariance features of the process. We analyze two types of stationary and dissipative $1$-dimensional scale-invariant processes: regularized fractional Brownian motion and regularized multifractal random walk. For both processes, the structure and dynamics of the embedding space are determined by their scale-invariant properties.
\end{abstract}


\maketitle

\section{Introduction}
\label{sec:intro}

In nature, plenty of phenomena such as turbulence~\cite{Frisch1995,Chevillard2012}, rainfall~\cite{Venugopal2006}, clouds~\cite{Freischem2024}, percolation~\cite{Rammal1983} or climate~\cite{Lovejoy2013}, present multiscale behaviors that can be modeled through scale-invariant stochastic processes~\cite{mandelbrotFractionalBrownianMotions1968,bacryMultifractalStationaryRandom2002,Robert2008,Chevillard2010}. These processes, that can be classified in monofractals and multifractals, have non-differentiable dynamics characterized by Holder-continuous time series~\cite{mandelbrotFractionalBrownianMotions1968}.

Given a time-series issued from the observation of a scale-invariant phenomenon, multifractal analysis consists in characterizing the range of displayed singularities $I=[h_{min},h_{max}]$ and their associated fractal dimensions, or said otherwise, the singularity spectrum $D(h)$. Multifractal analysis for stochastic processes includes multiscale methods based on structure functions~\cite{Muzy1993,GraneroBelinchon2025}, cumulants~\cite{castaingLogsimilarityTurbulentFlows1993,Chevillard2010}, wavelet coefficients~\cite{Muzy1993, Leonarduzzi2014} or information theory~\cite{GraneroBelinchon2018}.

The Multifractal formalism is also useful to characterize the scaling structure of strange attractors resulting from differentiable dynamical systems living in finite-dimensional phase spaces~\cite{Paladin1987}. In this case, multifractal analysis is performed through the measure of dimensions such as the capacity dimension $D_0$~\cite{Eckmann1985} or the correlation dimension $D_2$~\cite{Grassberger1983}. Of special interest are the generalized dimensions $D_q$~\cite{Balatoni1956,Badii1985,Arneodo1987,Riedi1995, Alber1998}, which are related to the singularity spectrum \textit{via} the Legendre transform~\cite{Arneodo1987,Paladin1987} and generalize the capacity and correlation dimensions. The most common estimators of the generalized dimensions are based on the study of the probability distribution of the scale-dependent local density in the phase space~\cite{Badii1985, Riedi1995, Alber1998, Paladin1987, Arneodo1987, Hadyn2002, Mantica2010}. New concepts to characterize the phase space of fractals such as the return times~\cite{Caby2019, Mantica2010} or the hitting times~\cite{Caby2019} also ground on measures of scale-dependent local densities and are linked to the generalized dimensions~\cite{Mantica2010, Hadyn2002, Caby2019}.

Extreme value theory was also used to describe the fractal properties of the phase space of deterministic and stochastic chaotic systems~\cite{Faranda2018, Caby2020, Caby2019, Alberti2023}. This characterization commonly grounds on the study of the probability distribution of the distances of neighbor states, or \emph{analogues}, with respect to a target state~\cite{Lucarini2012, Moloney2019, platzerProbabilityDistributionsAnalogToTarget2021, Alberti2023}. Studying the probability distribution of these distances allows to measure local dimensions in the phase space~\cite{platzerProbabilityDistributionsAnalogToTarget2021} that are informative of the correlation dimension~\cite{Faranda2018} and the generalized dimensions~\cite{Caby2019}. Moreover, studying the analogue's time evolution, allows to develop forecast methodologies~\cite{platzerUsingLocalDynamics2021c, lguensatAnalogDataAssimilation2017a} and characterizes the dynamics of the attractor~\cite{Paladin1987, Sapsis2013, GraneroBelinchon2026}. Several works have illustrated the main limitations and drawbacks of characterizing attractors with local density estimators, and more particularly, with local dimension estimators. First, these approaches are affected by the curse of dimensionality~\cite{Pons2020} and by uneven samplings of the phase space~\cite{Platzer2025a,Platzer2025b}. Second, they ground on the hypothesis of a regular variation of the probability measure which may be not adapted even for highly idealized dynamical systems~\cite{Amo2025}.

Inspired by these works, we propose to study scale-invariant stochastic processes in reconstructed $p$-dimensional spaces that are built from observed time series using Takens embedding~\cite{Takens1981,Abarbanel1992}. In an abuse of language, we will refer to these finite-dimensional reconstructed spaces as ``phase spaces''. We perform multifractal analysis by studying the structural and Lagrangian properties of analogue ensembles in these spaces. The \emph{analogues} of a given \emph{target state} in a finite-dimensional phase space are defined as its nearest neighbors with respect to a distance measure in this space.

First, we focus on the structural properties of the phase space by analyzing the volumes occupied by analogues in the different regions of the space. Previous methodologies characterize scale-dependent local densities by fixing volumes and measuring the number of analogue states within them~\cite{Badii1985,Riedi1995,Alber1998,Hadyn2002,Mantica2010}. This implies that the number of analogues within the volume is measured at a given scale of analysis that varies to perform multiscale analysis. Alternatively, we propose to fix a number $k$ of analogue states (or mass), such as in~\cite{Badii1985,Meisel1994,platzerProbabilityDistributionsAnalogToTarget2021}, and to measure the volume they occupy depending on the region of the phase space. The probability distribution of these volumes characterizes the distribution of significative scales in the process and is related to its multifractal properties. The volume linked to a target state is informative of the order of the singularity of this region of the phase space. Contrary to recent works that study the local probability distributions of analogue-to-target distances~\cite{platzerProbabilityDistributionsAnalogToTarget2021, Alberti2023}, we describe each local state by the volume occupied by its $k$-nearest neighbor states, and we study the probability distribution of these volumes in the full phase space, \textit{i.e.} each measured volume corresponds to a different target-state. 

Second, we study the Lagrangian properties of the analogue states in the phase space. More particularly, how the volume occupied by an ensemble of analogues evolves in time when these states disperse. Analysing the evolution of the volume occupied by the analogue's successors at different time-scales allow us to perform a multiscale analysis of the process dynamics.

We use both proposed approaches to characterize the multifractal properties of two stationary and dissipative scale-invariant stochastic processes: a regularized fractional Brownian motion (r-fBm), that is monofractal, and a regularized multifractal random walk (r-MRW), that is multifractal. These processes are specially interesting since they are commonly used to model long memory processes such as fluid turbulence~\cite{Chevillard2010, Pereira2016} or stock markets~\cite{Saakian2011,Sattarhoff2023}. Moreover, they are classically used as benchmarks in multifractal analysis since their multifractal properties are well known. We illustrate how the probability distribution of local volumes is determined by the multifractal properties of the studied processes. In particular, we quantify the impact of the range of displayed singularities $I$ over the probability distribution of local volumes, and we show that only the width of $I$ impacts the variance of the PDF of local densities, while both, the position and width, impact its mean. The study of the dispersion of successors in time allows us to perform a multiscale analysis of the dynamics of the states in the phase space. We found empirical relationships between the multifractal properties of the studied processes and the successors dispersion in the phase space. While the use of analogues is well established in dynamical systems and forecasting~\cite{Caby2020, platzerProbabilityDistributionsAnalogToTarget2021, lguensatAnalogDataAssimilation2017a}, and local-density estimators based on nearest-neighbors~\cite{Platzer2025a} share important similarities with to the proposed measure of local volumes, the relationship between analogue-successor dispersion and the multifractal properties of the process is novel.

In section~\ref{sec:multifractal}, we present two visions of multifractal analysis: one is based on the study of the probability distributions of the increments of the process, and the other is based on the characterization of the structural properties of a reconstructed finite-dimensional phase space. In this section, we also present the mathematical definition of the r-fBm and r-MRW. In section~\ref{sec:analogue}, we present the proposed framework for multifractal analysis: the phase space reconstruction, the definition of ensembles of analogues and successors and the measures of structural and dynamical characteristics. In section~\ref{sec:results}, we illustrate how the proposed framework describes the multifractal properties of the studied processes and in section~\ref{sec:conclusions}, we discuss the advantages and disadvantages of this framework as well as future perspectives.

\section{Multifractal formalism}
\label{sec:multifractal}

\subsection{Scale invariance}
\label{sec:selfsimil}

A 1-dimensional stochastic process $X=\{x(t)\}$, with increments $\delta_{\tau} X = \{x(t)-x(t-\tau)\}$, is scale-invariant if every statistical moment of its increments behaves as a power law of the increment's size:
 
\begin{equation}
\mathbb{E} \left(|\delta_{\tau} X|^q \right)\underset{\tau\rightarrow 0}\sim  k_q \tau^{\zeta(q)}
\end{equation}

\noindent where $\zeta(q)$ is the scaling function, $k_{q}$ are constants that depend on the order of the statistical moment and $\tau$ is the scale. The singularity spectrum $D(h)$ is related to the scaling function by the Legendre transform:

\begin{equation}
    \zeta(q) = \min_{h}(qh + d - D(h))
\end{equation}

\noindent where $d=1$ is the dimension of the process~\cite{Chevillard2012}. 

Scale-invariant processes can be distinguished into two main families:

\begin{itemize}

\item Monofractal processes present a linear scaling function $\zeta(q)=q{\cal H}$, with the Hurst exponent ${\cal H}$ characterizing the roughness of the process. The singularity spectrum is single-valued $D(h)=\delta(h-\mathcal{H})$. These processes do not exhibit intermittency, consequently they don't have small-scale extreme events and the shape of the PDF of the increments does not evolve across scales.

\item Multifractal processes have a non-linear scaling function. In this work, we will focus on log-normal multifractal models whose scaling function is of the form $\zeta(q)=q{\cal H}-\frac{c_2}{2}q^2$. The singularity spectrum has a parabolic shape with $\mathcal{H}$ characterizing its maximum and $c_2$ its width. Consequently, the most common roughness is characterized by $\mathcal{H}$, and $c_2>0$ is the intermittency coefficient that characterizes the width of the range of the existing singularities. These processes present intermittency: the shape of the PDF of their increments evolves across scales and presents extreme events at small scales.
\end{itemize}

\subsection{Multifractal objects in finite-dimensional phase spaces}
\label{sec:multiphase}

The spectrum of generalized dimension of the stochastic process $X=\{x(t)\}$ can be characterized through scale-dependent local density measures in a reconstructed $p$-dimensional phase space~\cite{Paladin1987, Hadyn2002}. Considering a discrete realization of the stochastic process, the percentage of states $x(t')$ within a $p$-dimensional ball of radius $l$ centered at the state $x(t)$ is: 

\begin{equation}
    n_t(l) = \lim_{N\to \infty} \frac{1}{N-1}\sum_{t \neq t'} \theta(l-|x(t)-x(t')|)
\end{equation}

\noindent where $N$ is the number of samples in the realization and $\theta$ is the Heaviside step function. Since the volume of the ball is fixed, the number of states within the ball (its mass) can be interpreted as a measure of local density at the scale $l$.

The probability distribution of this scale-dependent densities fully characterizes the fractal object. In particular, generalized dimensions can be defined from:

\begin{equation}
    \left\langle n_t(l)^q \right\rangle_t \propto l^{q D_{q+1}}
\end{equation}

\noindent where $\left\langle  \right\rangle_t$ is the average in time.

The capacity dimension and the correlation dimension correspond to the generalized dimensions $D_0$ and $D_2$ respectively. More importantly, the generalized dimensions and the singularity spectrum are related by the Legendre transform~\cite{Paladin1987}. Consequently, multifractal analysis can be performed through the characterization of the probability distribution of the local densities in a $p$-dimensional phase space.

In this section, we provided only a simple introduction of generalized dimensions to illustrate the connection between the multifractal analysis performed on the increments of the stochastic process and the multifractal analysis performed on reconstructed phase spaces, see~\cite{Caby2019} and references therein for more details.

\subsection{Multifractal stochastic processes}
\label{sec:processes}

We study two types of $1$-dimensional stochastic processes, one monofractal and the other multifractal, called respectively regularized fractional Brownian motion and regularized Multifractal Random Walk~\cite{Robert2008, Chevillard2010, Pereira2016}. Both processes are modeled with the following stochastic integral~\cite{Robert2008, Pereira2016}:

\begin{equation}
X_{\mathcal{H},c_2,\tau_K}(t) = \int_{\mathbb{R}} \psi_T(t-t^\prime) P_{\mathcal{H},\tau_K} (t-t^\prime) M_{c_2,\tau_K}(t^\prime)W(t^\prime)d t^\prime
\label{def:field}
\end{equation}

\noindent where $t$ and $t^\prime$ denote $1$-dimensional time vectors, $W(t)$ is a Gaussian white noise, $0<\mathcal{H}<1$ is the Hurst exponent and $c_2$ is the intermittency coefficient. The term $\psi_T(t)$ is a Gaussian large-scale cut-off~\cite{Chevillard2010}. The term $P_{\mathcal{H},\tau_K}(t) = \frac{1}{||t||_{\tau_K}^{1/2-\mathcal{H}}}$, is a kernel providing a power spectrum with a power law behavior of exponent $2\mathcal{H}+1$. The norm $||t||_{\tau_K} = \sqrt{||t||^2+\tau_K^2}$ in the denominator is a regularized $L^2$-norm with $\tau_K>0$ being the regularization scale and $||.||$ the $L^2$-norm.

The term $M_{c_2,\tau_K}(t)$ is a multiplicative chaos \cite{Robert2008} defined as:

\begin{equation}
M_{c_2,\tau_K}(t)=e^{-\sqrt{c_2} X_{\tau_K}(t)-c_2\mathbb{E}\{X_{\tau_K}^2(t)\}}
\end{equation}
\noindent where $X_{\tau_K}(t)$ is a log-correlated Gaussian noise with autocovariance function:

\begin{equation}\label{eq:autocov}
\mathbb{E} \{X_{\tau_K}(t)X_{\tau_K}(t^\prime) \}  \underset{||t-t^\prime||_{\tau_K}\rightarrow 0}\sim -\log(||t-t^\prime||_{\tau_K})
\end{equation}

The process $X_{\mathcal{H},c_2,\tau_K}(t)$ presents three different domains of scales. A dissipative domain at scales smaller than $\tau_K$, an inertial domain for scales larger than $\tau_K$ and smaller than $T$ and an integral domain at scales larger than $T$. The statistical moments of the increments of $X_{\mathcal{H},c_2,\tau_K}(t)$ behave differently depending on the domain of scales:

\begin{numcases}{\mathbb{E} \left(|\delta_{\tau} X_{\mathcal{H},c_2,\tau_K}|^q \right) \sim }
    \tau^{q} &for  $\tau<\tau_K$ \label{eq:Sdis}\\
    \tau^{\zeta(q)} &for $\tau_K<\tau<T$ \label{eq:Siner}\\
    \left(2 \left\langle X_{\mathcal{H},c_2,\tau_K}^{2} \right\rangle \right)^{q/2} &for $\tau > T$ \label{eq:Sint}
\end{numcases}

\noindent Consequently, this process exhibits scale-invariance in the inertial domain.

The regularized fractional Brownian motion corresponds to the case $c_2=0$. In this case, $X_{\mathcal{H},c_2=0,\tau_K}(t)$ is Gaussian, monofractal, stationary and its statistical properties are fully defined by its Hurst exponent $\mathcal{H}$~\cite{Chevillard2012}. This process does not display intermittency and consequently the probability distributions of its increments remain Gaussian across scales \textit{i.e.} small scales do not present extreme events.

The regularized multifractal random walk corresponds to the case $c_2 > 0$. This stochastic process, $X_{\mathcal{H},c_2,\tau_K}(t)$, is multifractal and stationary, its large scales have Gaussian probability distributions while the small scales present heavy tailed ones. The statistics of r-MRW are prescribed by the log-normal multifractal model and so they are fully defined by its Hurst exponent and intermittency coefficient.  

The algorithms used to generate these multiscale stochastic processes are freely available at: https://gitlab.com/sroux67/multivariate-multifractal-field-synthesis.

\section{Analogue ensembles}
\label{sec:analogue}

\subsection{Phase space reconstruction and analogues definition}
\label{sec:anadef}

Given a time series $X=\{x(t)\}$ sampled homogeneously at time intervals $dt$, we use Takens embedding~\cite{Takens1981, Eckmann1985, Abarbanel1992} to reconstruct a $p$-dimensional phase space where the state at time $t$ is:

\begin{equation}
    \vec{x}^{(p)}(t) = 
    \begin{pmatrix}
        x(t) \\
        x(t-m \,dt) \\
        \vdots \\
        x(t-(p-1)\, m\,dt)
    \end{pmatrix}
\end{equation}

Takens embedding depends on two parameters: the dimension of the phase space $p$ and the time scale $m \, dt$.

In this $p$-dimensional reconstructed phase space, two states $\vec{x}^{(p)}(t)$ and $\vec{x}^{(p)}(t')$ at two different times $t$ and $t'$ are \emph{analogues} with respect to a given distance definition $J$, if $J(\vec{x}^{(p)}(t), \vec{x}^{(p)}(t')) < \epsilon$, where $\epsilon$ is a threshold. 

Given a target state $\vec{x}^{(p)}(t)$, the ensemble of $k$ closest analogues $\textbf{x}_{\textbf{a}}(t) = \left\lbrace \vec{x}^{(p)}(t'_i) \right\rbrace_{1\leq i \leq k}$ is composed of the $k$ closest nearest neighbors of the state. This definition of the analogues ensemble implies a threshold $\epsilon_{t'_k}$ defined as the distance between the observed state and its $k$-th nearest neighbor, $\epsilon_{t'_k}=J(\vec{x}^{(p)}(t),\vec{x}^{(p)}(t'_k))$. Consequently, the threshold depends on the observed state, and thus on the sampled region of the phase space. Each analogue $\vec{x}^{(p)}(t'_i)$ has an associated successor $\vec{x}^{(p)}(t'_i+\tau)$ at time $\tau$. We define the ensemble of analogue successors at time $\tau$ as $\textbf{x}_{\textbf{s}}(t+\tau) = \left\lbrace \vec{x}^{(p)}(t'_i+\tau) \right\rbrace_{1\leq i \leq k}$. 

This analogues definition is based on two steps: 1) the reconstruction of a finite-dimensional phase space from a time series and 2) the search for the nearest neighbors of the observed target states. For the reconstruction, we consider Takens embeding with $m=1$ ($m\,dt=dt$) and $p=3$. Appendix \ref{sec:appendix} contains a sensitivity analysis on the Takens embedding parameters of the reconstruction as well as a discussion on the choice of these parameters. For the analogues search, we consider the Euclidean distance, as suggested by~\cite{platzerUsingLocalDynamics2021c,lguensatAnalogDataAssimilation2017a}. Different phase space reconstructions and distance definitions can be used~\cite{Takens1981, Sutulovic2025, Platzer2025}.

In practice, we construct a historical database used to search for analogues, along with an independent realization that defines the target states in the phase space. We refer to these as the \emph{database} and the \emph{measure}, respectively. Throughout the article, $N_D$ denotes the number of samples in the database, while $N_M$ represents the number of samples in the measure. To ensure that the $p$-dimensional phase space is densely sampled, and thus that the $k$ analogues lie close enough to each target state, the database must be sufficiently large. Moreover, as the dimension of the phase space increases, an even larger database is required to obtain meaningful analogues for each target state~\cite{Cecconi2012, Chibbaro2014}. Finally, the measure realization must be long enough to adequately cover the entire phase space so that the resulting statistics are representative.

\subsection{Volume occupied by analogues and successors}
\label{sec:density}

For each target state in the measure realization, we define the volume occupied by the analogues as the pairwise Euclidean distance between the $k$ states of the ensemble:

\begin{equation}\label{eq:errordefana}
\delta_a(t) = \frac{2}{k(k-1)}\sum_{i=2}^{k}\sum_{j<i} \left(\vec{x}^{(p)}(t'_i)-\vec{x}^{(p)}(t'_j) \right)^2  
\end{equation}

\noindent which can be interpreted as a measure of the rarity of the observed state $\vec{x}^{(p)}(t)$.

Several works used the distance to the best neighbor~\cite{Badii1984,Badii1985,Meisel1994,Kostelich1989}, and more recently the distance to the $k$-th best analogue~\cite{platzerProbabilityDistributionsAnalogToTarget2021,lguensatAnalogDataAssimilation2017a} to define the characteristic local scale. However, these estimators based on a single distance to a neighbor are statistically less robust than the proposed average and lead to estimations with very high variance when applied to the multifractal stochastic processes presented in section~\ref{sec:processes}.

Equivalently, we can define the volume occupied by the successors after a delay $\tau$ as the pairwise Euclidean distance between the $k$ analogue successors:

\begin{equation}\label{eq:errordefsuc}
\delta_s(t+\tau) = \frac{2}{k(k-1)}\sum_{i=2}^{k}\sum_{j<i} \left(\vec{x}^{(p)}(t'_i+\tau)-\vec{x}^{(p)}(t'_j+\tau) \right)^2    
\end{equation}

\noindent which studied in function of $\tau$ informs about the dynamics of the system in the reconstructed phase space.

For a measure realization with $N_M$ states, we will measure $N_M$ volumes $\delta_a$ and $\delta_s$, one for each target state of the measure. The statistics will be performed on the ensemble of these states.

\section{Results}
\label{sec:results}

We study five regularized fractional Brownian motions characterized by their Hurst exponents $\mathcal{H} \in \left\lbrace 0.3,0.4,0.5,0.6,0.7 \right\rbrace$. We also studied regularized multifractal random walks with $\mathcal{H} \in \left\lbrace 0.3,0.4,0.5,0.6,0.7 \right\rbrace$ and $c_2 \in \left\lbrace 0.025,0.05,0.075,0.1 \right\rbrace$. For all processes, the regularization and integral scales are respectively fixed to $\tau_K = 5 dt$ and $T = 2350 dt$. In our experiment, the database where looking for analogues contains $N_D=5 \times 2^{21}$ samples, corresponding to approximately $4460$ integral scales $T$. This ensures a big enough database to consider that the phase space is densely sampled. A sensitivity analysis on the size of this database is done in appendix \ref{sec:appendix}. The measure realization contains $N_M=2^{21}$ samples covering the full phase space. We generate ensembles with $k=50$ members. The values of $T$, $\tau_K$, $N_D$ and $N_M$ have been fixed as in~\cite{GraneroBelinchon2026} where a experimental turbulent velocity time series was studied with a similar methodology. Different $k$ values between $k=50$ and $k=200$ were tested without qualitative changes in the results.

\subsection{Statistics of the analogues' volume}
\label{sec:res1}

For each specific target state in the measure, we quantify the volume occupied by its analogues. Then, we study the structural properties of the phase space through the probability distribution of the logarithm of the volumes occupied by the ensembles of analogues $\mathbb{P} \left(\log(\delta_a)\right)$. In particular, we focus on the impact of $\mathcal{H}$ and $c_2$ on the mean and standard deviation of $\mathbb{P} \left(\log(\delta_a)\right)$. Each measured volume of the probability distribution corresponds to a different target-state.

\begin{figure}[ht]
 \includegraphics[width=0.9\linewidth]{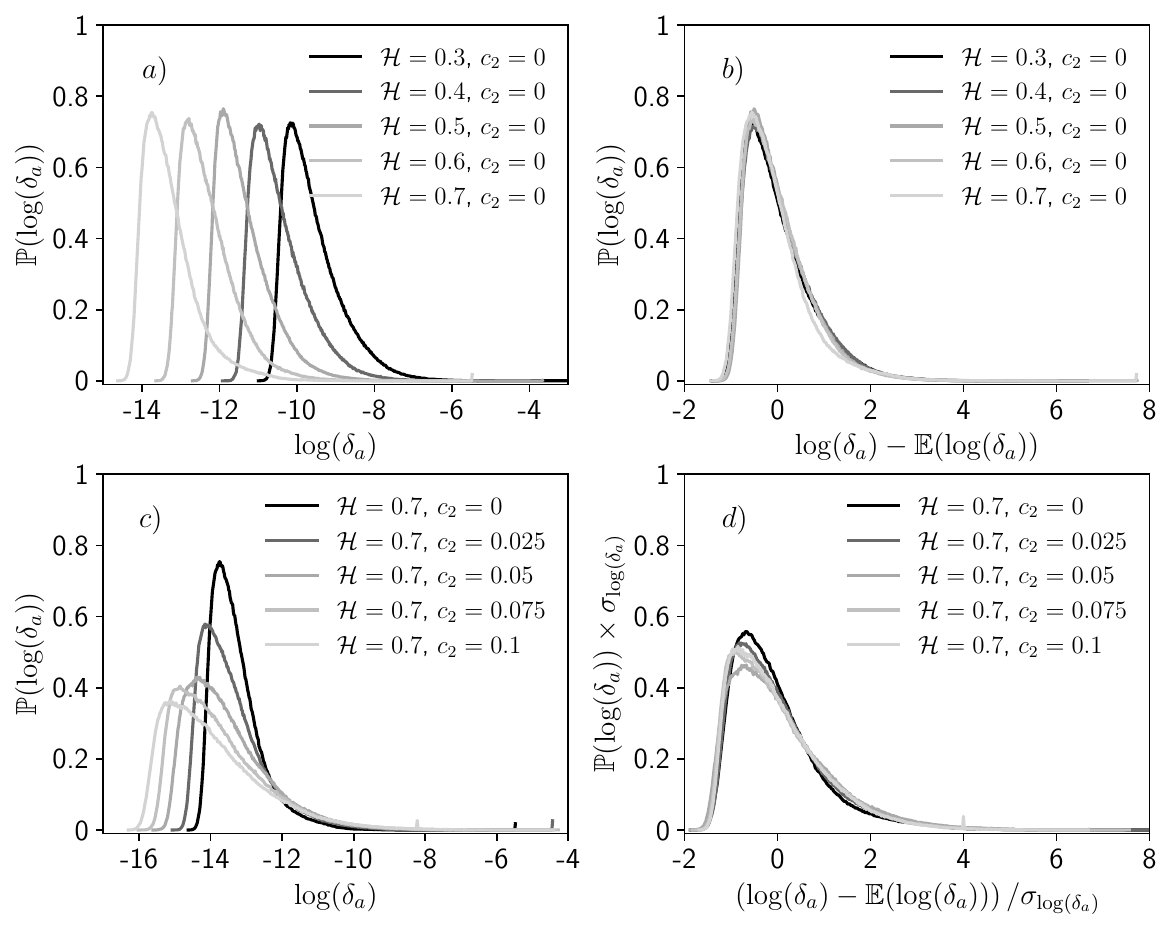}
 \centering
 \caption{a) and c) Probability distribution of the logarithm of the volume occupied by the analogues $\mathbb{P}(\log(\delta_a))$ in function of $\log(\delta_a)$. b) Probability distribution of the logarithm of the volume occupied by the analogues $\mathbb{P}(\log(\delta_a))$ in function of $\log(\delta_a) - \mathbb{E}(\log(\delta_a))$. d) Probability distribution of the logarithm of the volume occupied by the analogues scaled by its standard deviation $\mathbb{P}(\log(\delta_a)) \times \sigma_{\log(\delta_a)}$ in function of $\left( \log(\delta_a) - \mathbb{E}(\log(\delta_a)) \right)/\sigma_{\log(\delta_a)}$. In a) and b) for regularized fractional Brownian motions with $\mathcal{H} \in \left\lbrace 0.3,0.4,0.5,0.6,0.7\right\rbrace$ and $c_2=0$. In c) and d) for regularized multifractal random walks with $\mathcal{H}=0.7$ and $c_2 \in \left\lbrace 0,0.025,0.05,0.075,0.1\right\rbrace$.}\label{fig4}
\end{figure}

Figures~\ref{fig4} a) and c) show the probability distribution $\mathbb{P} \left(\log(\delta_a)\right)$ in function of $\log(\delta_a)$ respectively for different $\mathcal{H}$ and $c_2=0$ and for different $c_2$ and $\mathcal{H}=0.7$. Figures~\ref{fig4} b) and d) are rescaled versions allowing to collapse all the probability distributions. 

\begin{figure}[ht]
 \includegraphics[width=1\linewidth]{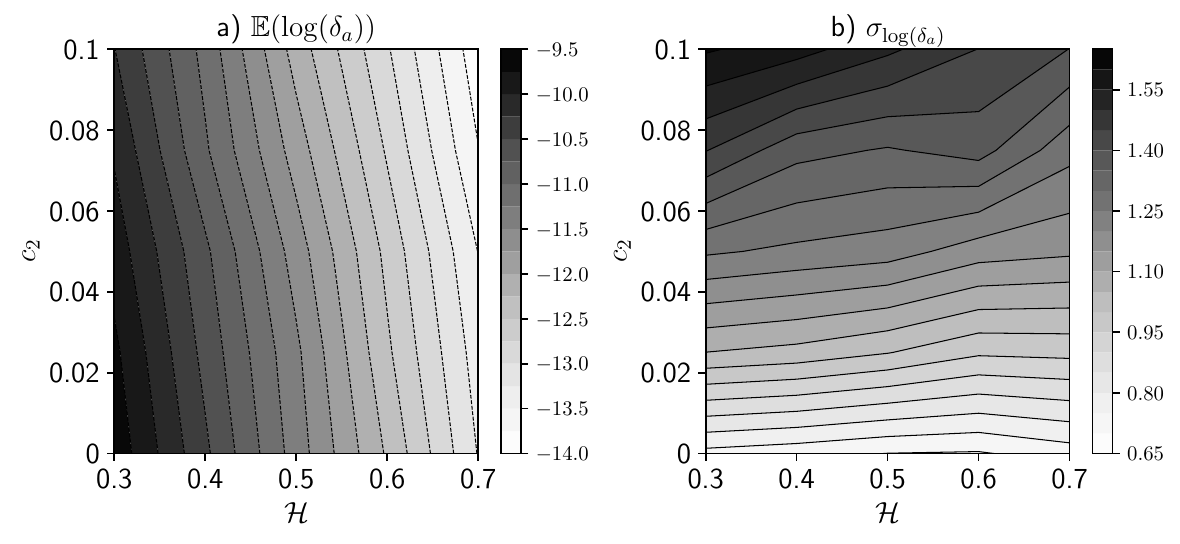}
 \centering
 \caption{a) Mean $\mathbb{E}(\log(\delta_a)$ and b) standard deviation $\sigma_{\log(\delta_a)}$ of the logarithm of the volume occupied by the analogues in function of the Hurst exponent $\mathcal{H}$ and the intermittency coefficient $c_2$.}\label{fig5}
\end{figure}

In the case of r-fBm, changes in the Hurst exponent only impact the mean of the probability distribution $\mathbb{P} \left(\log(\delta_a)\right)$ while the standard deviation remains unchanged. In the case of r-MRW, changes in $c_2$ impact both the mean and standard deviation of $\mathbb{P} \left(\log(\delta_a)\right)$. This is better observed in figures~\ref{fig5} a) and b), which illustrate respectively the behavior of the mean $\mathbb{E}(\log(\delta_a)$ and standard deviation $\sigma_{\log(\delta_a)}$ of the logarithm of the volume occupied by the analogues in function of the values of $\mathcal{H}$ and $c_2$. The probability distribution of the volume occupied by the analogues $\mathbb{P} \left(\log(\delta_a)\right)$ is influenced by the Hurst exponent and the intermittency coefficient characterizing each process. Thus, the roughness of the stochastic process defines the probability distribution of $\delta_a$, which can be interpreted as a characterization of the distribution of the local volumes, and of the quality of the analogues for each state. On the one hand, a large value of $\mathbb{E}(\log(\delta_a)$ implies larger volumes occupied by the analogues and so bad qualities of the found analogues. On the other hand, large values of $\sigma_{\log(\delta_a)}$ indicate a high heterogeneity in the volumes occupied by the analogues and consequently a high heterogeneity of the quality of the analogues. 

\subsection{Statistics of the analogue-successor's volume}
\label{sec:res2}

In this section, we study how the roughness of the process impacts the dynamics of the analogues, and more particularly how the dependence of $\delta_s(t+\tau)$ on $\tau$ and on $\delta_a(t)$ is determined by the roughness of the process. In this way, we study the analogues' dynamics and not only the static probability distribution of their volumes.

In~\cite{GraneroBelinchon2026}, we proposed that for scale-invariant processes 1) the mean of the volume of the successors behaves as a power law of the time $\tau$ and 2) the volume of the successors behaves as a power law of the volume of the analogues:

\begin{eqnarray}
    \left\langle \delta_s(t+\tau)\right\rangle &\propto& \tau^{\beta} \label{eq:propos}\\
    \delta_s(t+\tau) &\propto& (\delta_a(t))^{\alpha(\tau)} \label{eq:propos2}
\end{eqnarray}

\noindent where $\left\langle \, \right\rangle$ is the average over the collection of target-states. Here, we study the dependence of $\beta$ and $\alpha$ on $\mathcal{H}$ and $c_2$.

\begin{figure}[ht]
 \includegraphics[width=0.5\linewidth]{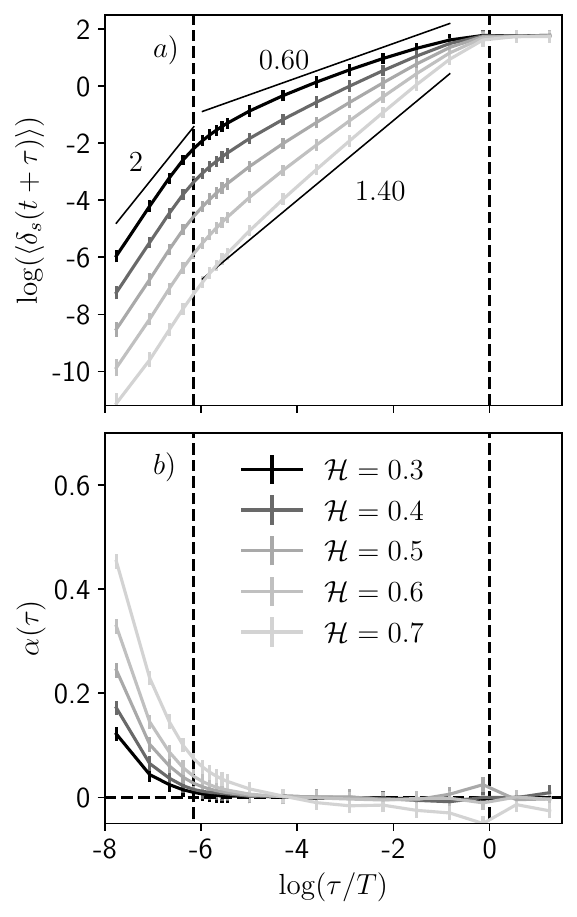}
 \centering
 \caption{a) Logarithm of the mean volume occupied by the successors $\log(\delta_s(t+\tau))$ and b) exponent $\alpha(\tau)$, both in function of the logarithm of the time scale $\log(\tau/T)$ for regularized fractional Brownian motions with $\mathcal{H} \in \left\lbrace 0.3,0.4,0.5,0.6,0.7\right\rbrace$ and $c_2=0$.}\label{fig1}
\end{figure}

Figure~\ref{fig1} shows $\left\langle \delta_s(t+\tau)\right\rangle$ and $\alpha(\tau)$ in function of $\log(\tau/T)$ for the different studied r-fBm. The logarithm of the mean volume of the successors presents three different behaviors corresponding to the three different domains of scales:

\begin{numcases}{\log(\left\langle \delta_s(t+\tau)\right\rangle) \sim }
    2 \log(\tau/T) &for  $\tau<\tau_K$ \label{eq:ddis}\\
    2 \mathcal{H} \log(\tau/T) &for $\tau_K<\tau<T$ \label{eq:diner}\\
    \log(2p) &for $\tau > T$ \label{eq:dint} 
\end{numcases}

In the dissipative domain, the mean volume of the successors behaves as $2\log(\tau/T)$ independently of the Hurst exponent $\mathcal{H}$. In the inertial domain, it behaves linearly in $\log(\tau/T)$ with the slope being $\mathcal{H}$. In the integral domain, all the curves collapse to a plateau that depends on the dimension of the reconstructed phase space. The exponent $\alpha$ is equal to zero in the integral and inertial domains. In the dissipative domain, it takes positive values that increase with the Hurst exponent.

\begin{figure}[ht]
 \includegraphics[width=0.8\linewidth]{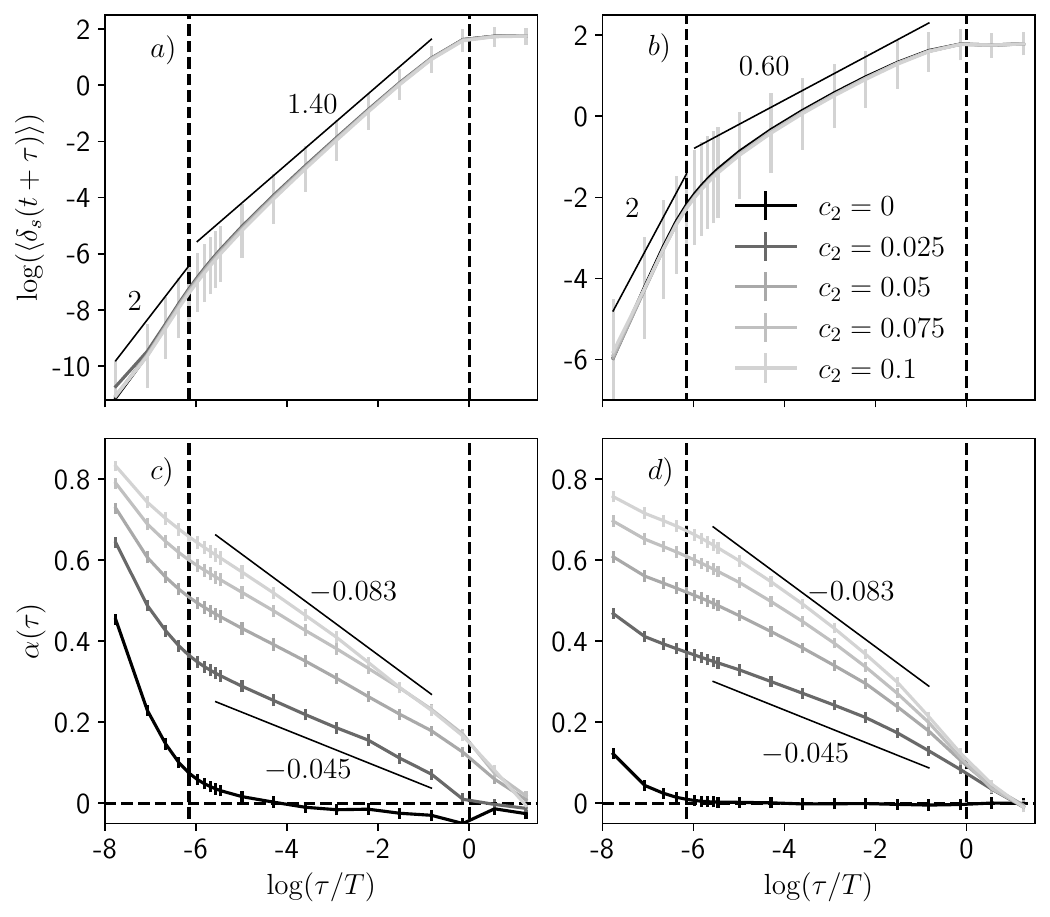}
 \centering
 \caption{Logarithm of the mean volume occupied by the successors $\log(\delta_s(t+\tau))$, in a) and b), and exponent $\alpha(\tau)$, in c) and d), in function of the logarithm of the time scale $\log(\tau/T)$ for regularized multifractal random walks. In a) and c) with $\mathcal{H} = 0.7$ and in b) and d) with $\mathcal{H} = 0.3$. The intermittency coefficient is $c_2 \in \left\lbrace 0,0.025,0.05,0.075,0.1\right\rbrace$.}\label{fig2}
\end{figure}

Figure~\ref{fig2} shows $\left\langle \delta_s(t+\tau)\right\rangle$ and $\alpha(\tau)$ in function of $\log(\tau/T)$ for different studied r-MRW. In a) and c) the r-MRW has $\mathcal{H}=0.7$ and in b) and d) $\mathcal{H}=0.3$. Independently of the processes, $\left\langle \delta_s(t+\tau)\right\rangle$ follows equations (\ref{eq:ddis}), (\ref{eq:diner}) and (\ref{eq:dint}), having a different scaling behavior for each domain of scales. The exponent $\alpha$ is equal to zero for scales in the integral domain, and it increases when the scale decreases for scales in the inertial and dissipative domains. The behavior of $\alpha$ in the inertial domain seems to be independent of $\mathcal{H}$ and, for $p=3$, we found an empirical relationship on $c_2$ of the form:

\begin{equation}
    \alpha(\tau) \sim -0.27 \sqrt{c_2} \log(\tau/T) \,\,\, \text{for} \,\,\, \tau_K<\tau<T . \label{eq:alpha}
\end{equation}

All the processes have an inertial domain of scales where $\left\langle \delta_s(t+\tau)\right\rangle$ follows the scaling behavior proposed in (\ref{eq:propos}) with $\beta=\mathcal{H}$, and $\delta_s(t+\tau)$ follows the scaling behavior in (\ref{eq:propos2}) with $\alpha(\tau)$ following (\ref{eq:alpha}).

\section{Discussion and conclusions}
\label{sec:conclusions}

We proposed a framework for analyzing the scale-invariant properties of stochastic processes in reconstructed finite-dimensional phase spaces. In this work, we used Takens embedding to reconstruct these phase spaces. However, alternative reconstruction methodologies~\cite{Ouala2020,Sutulovic2025} should be investigated in the future. We studied two types of scale-invariant processes: r-fBm which is monofractal and whose roughness is fully defined by the Hurst exponent $\mathcal{H}$, and r-MRW which is multifractal and its roughness is prescribed by $\mathcal{H}$ and the intermittency coefficient $c_2$. We focused on the statistical description of the structural and dynamical properties of the reconstructed space of the process and their link with the roughness of the process. More particularly, we characterized the influence of $\mathcal{H}$ and $c_2$ on 1) the probability distribution of the logarithm of the volume occupied by the analogues $\mathbb{P} \left(\log(\delta_a)\right)$, 2) the time-evolution of the volume occupied by the successors $\delta_s(t+\tau)$ and 3) the dependance of $\delta_s(t+\tau)$ on the initial volume $\delta_a(t)$. On the one hand, we showed that both $\mathcal{H}$ and $c_2$ influence the mean of $\mathbb{P} \left(\log(\delta_a)\right)$, while only $c_2$ impacts its variance. On the other hand, we observed that, for time-scales in the inertial domain, the scaling of $\left\langle\delta_s(t+\tau)\right\rangle$ is characterized by $\mathcal{H}$, see equation (\ref{eq:diner}), while the dependence of $\delta_s(t+\tau)$ on $\delta_a(t)$ only relies on $c_2$ through equations (\ref{eq:propos2}) and (\ref{eq:alpha}).

The volume occupied by the $k$-nearest analogues of an observed state, as defined in (\ref{eq:errordefana}), is a measure of the rarity of the state: large volumes implies rare states. This is just because rare states live in regions of the phase space that are not frequently visited by the process and so for those states the $k$-nearest neighbors are further than usual. In this sense, large $\mathbb{E} \left(\log(\delta_a)\right)$ implies that the phase space is not densely sampled. We found that the higher the Hurst exponent the smaller $\mathbb{E} \left(\log(\delta_a)\right)$, which indicates that smoother processes occupy the phase space more densely, while rougher processes tend to be sparser. The variance of $\mathbb{P} \left(\log(\delta_a)\right)$ describes the heterogeneity of the phase space sampling. Small variances indicate that the volume occupied by the analogues depends barely on the observed state, and so, this corresponds to a homogeneous sampling where there are not, or not too many, extreme events. On the contrary, large variances indicate that the volume occupied by the analogues depends a lot on the state, and denote the existence of rare events. We observe, as expected, that a higher $c_2$ leads to a higher variance of $\mathbb{P} \left(\log(\delta_a)\right)$.

The dynamics of the analogues in the reconstructed phase space are also governed by the roughness properties of the multifractal process. We observe in (\ref{eq:dint}) that the volume occupied by the successors in function of the time $\tau$ converges to a fixed value for $\tau>T$. This value should be interpreted as the total volume occupied by the process in this reconstructed phase space. In the inertial domain of scales, where the process is scale-invariant, the mean volume occupied by the successors increases in time following the power law behavior prescribed in (\ref{eq:diner}). The higher the Hurst exponent the faster the dispersion of the successors, \textit{i.e.} smoother processes disperse faster than rougher ones. Very interestingly, at scales in the dissipative domain, $\tau<\tau_K$, the volume occupied by the successors depends on the volume occupied by the analogues for both studied processes r-fBm and r-MRW. This is not the case for scales in the inertial domain where this dependence remains only in the case of r-MRW. In the domain of scales where the studied processes are scale-invariant, only multifractal processes present a dependence of $\delta_s(t+\tau)$ on $\delta_a(t)$, highlighting the main role of $c_2$ in the dynamics of the analogues. The existence of a variety of singularities in multifractal processes leads to a variety of analogue's dynamics.

We proposed a framework for multifractal analysis based on the reconstruction of a phase space and the study of 1) the probability distribution of the volumes occupied by ensembles of analogues and 2) the Lagrangian properties of these ensembles. Contrary to classical approaches that compute statistics by varying the database where searching for analogues~\cite{platzerProbabilityDistributionsAnalogToTarget2021}, we made statistics on the ensemble of target-states for a fixed database. These statistics characterize the structural and Lagrangian properties of the full phase space. We studied how these properties depend on the Hurst exponent and the multifractal coefficient for r-fBm and r-MRW. These processes have respectively linear and parabolic scaling functions $\zeta(q)$. However, the proposed framework can be a good hint to characterize higher-order coefficients in scaling functions described by higher-order polynomials.

The quality of the analogues strongly depends on the size of the database and the dimension of the studied system~\cite{Cecconi2012} making the application of this kind of methodology in real-world high-dimensional systems not straightforward. However, several works already used analogue methods to study real-world high-dimensional systems~\cite{Platzer2025b, Faranda2026}. Moreover, very huge (of the order of the ones used in this paper) experimental databases of fluid dynamics already exist~\cite{Kahalerras1998, Chanal2000, Cheminet2022} and the presented methodology has been already used on them to study turbulence~\cite{GraneroBelinchon2026}. Applying this methodology to characterize atmospheric and oceanic dynamics is a promising direction for future research.

\appendix

\section{Sensitivity analysis on Takens embedding parameters and size of the analogues' database} \label{sec:appendix}

The proposed reconstruction of a finite-dimensional space is based on Takens embedding, that depends on two parameters: the embedding dimension $p$ and the embedding delay $m$. To study the impact of these parameters on the relationship between the dispersion of analogues and the multifractal features of the stochastic process, we directly analyse the impact of $p$ and $m$ on the estimation of $\mathcal{H}$ and $c_2$ for a r-MRW with $\mathcal{H}=0.3$ and $c_2=0.025$. 
Very importantly, $\log\left( \left\langle \delta_s(t+\tau)\right\rangle \right)$ and $\alpha(\tau)$ have a linear behavior in function of $\log(\tau/T)$ for a range of scales in the inertial domain. The slopes of these linear behaviors are measures of $\mathcal{H}$ and $c_2$ respectively. The region of fit is slightly altered for the different $p$ and $m$ combinations, but lies always in the inertial domain of scales. The empirical definitions of $c_2$ and $\mathcal{H}$ in function of the measured slopes that we use in this appendix are:

\begin{eqnarray}
    \mathcal{H} &= \frac{1}{2} \frac{d \log(\left\lbrace \delta_s(\tau)\right\rbrace)}{d\log(\tau/T)} \\
    c_2 &= (-\frac{4}{\log(p)} \frac{d \alpha(\tau)}{d\log(\tau/T)})^{2}
\end{eqnarray}

\begin{figure}[ht]
 \includegraphics[width=0.8\linewidth]{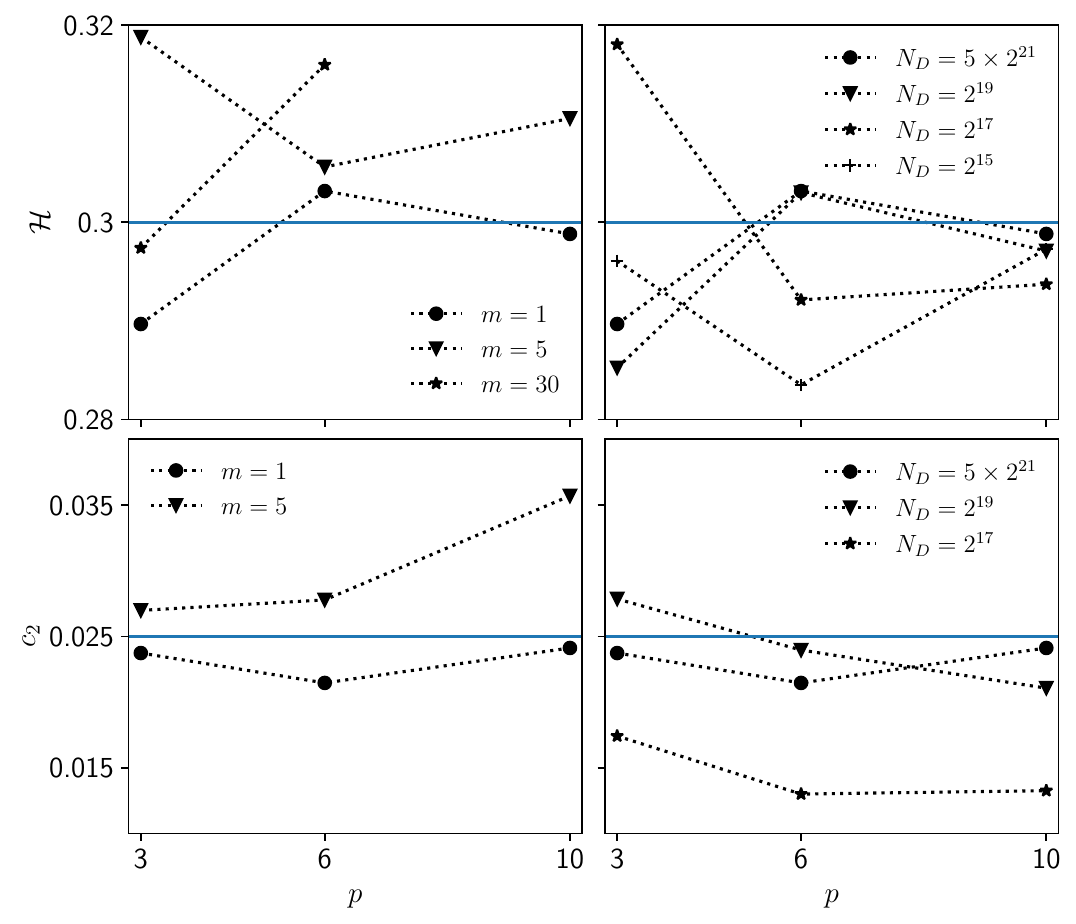}
 \centering
 \caption{Estimated Hurst exponent $\mathcal{H}$, a) and b), and intermittency coefficient $c_2$, c) and d), in function of the embedding dimension $p$. For different embedding delays $m$, a) and c), and different sizes of the database $N_D$, b) and d). The blue horizontal lines in a) and b) correspond to the real value of $\mathcal{H}$ and in c) and d) to the real value of $c_2$. The estimation of $\mathcal{H}$ was impossible for $p=10$ and $m=30$. The estimation of $c_2$ was impossible for $m=30$ and for $N_D=2^{15}$, in both cases independently of $p$.}\label{fig:app}
\end{figure}

Figure~\ref{fig:app} a) and c) illustrate the dependance of $\mathcal{H}$ and $c_2$ on the embedding dimension $p=[3,6,10]$ and for different embedding delays $m=[1dt, \tau_K=5dt, 30dt]$. These embedding delays correspond respectively to the smallest available time-scale, the regularization scale $\tau_K$ and a scale in the inertial domain, where the stochastic processes present scale-invariance. The error in $\mathcal{H}$ is smaller than $0.02$ ($6 \%$ of $\mathcal{H}$) for any pair of $p$ and $m$, except for $p=10$ and $m=30$ for which $\log\left( \left\langle \delta_s(t+\tau)\right\rangle \right)$ does not behave linearly in $\log(\tau/T)$ and so, it is not possible to estimate $\mathcal{H}$. The estimation of $c_2$ is more difficult with errors smaller than $0.005$ ($20 \%$ of $c_2$) for $m=1$ independently of $p$ and for $m=5$ when $p=[1,3]$. For $p=10$ and $m=5$, we estimate $c_2=0.035$. For $m=30$, independently of $p$, $\alpha(\tau)$ does not have a linear behavior in function of $\log(\tau/T)$ and so, it is not possible to estimate $c_2$.

We also studied the impact that the size $N_D$ of the database has on the estimation of $\mathcal{H}$ and $c_2$ for a r-MRW with $\mathcal{H}=0.3$ and $c_2=0.025$. Figure~\ref{fig:app} b) and d) show respectively the impact of $N_D$ on the estimation of $\mathcal{H}$ and $c_2$. We studied four different sizes of the database $N_D=[5 \times 2^{21},2^{19},2^{17},2^{15}]$ with different $p=[3,6,10]$ and fixed $m=1$. On the one hand, $\mathcal{H}$ is correctly estimated independently of $N_D$ and $p$. On the other hand, the estimation of $c_2$ is good and only slighly dependent of $p$ for $N=5 \times 2^{21}$ and $N_D=2^{19}$, but becomes worser when $N_D$ decreases. In particular when $N_D=2^{15}$, there is no linear behavior of $\alpha(\tau)$ on $\log(\tau/T)$ and so no possibility of measuring $c_2$.

The estimations are better for higher sizes of the database $N_D$, but with a very small difference between $N_D=2^{19}$ and $N_D=5\times2^{21}$. The smallest embedding delay $m=1$ provides a better characterization of the multifractal features, with only small differences between $m=1$ and $m=5$. However, for embedding delays corresponding to scales in the inertial domain, the estimations deteriorate. The impact of $p$ seems to be negligible when $m=1$ and large $N_D$ but becomes important when $m$ increases. Consequently, in this paper we fixed $m=1$, $p=3$ and $N_D=5\times2^{21}$.

\section*{Acknowledgment}

The author wishes to thank S. G Roux for providing the code used for the generation of scale-invariant stochastic processes. The author wishes to thank B. Dubrulle and P. Platzer for stimulating discussions. 




\bibliographystyle{elsarticle-num}
\bibliography{biblio}







\end{document}